\documentstyle[prl,aps,floats,epsf]{revtex}
\begin{document}
\draft

\twocolumn[\hsize\textwidth\columnwidth\hsize\csname @twocolumnfalse\endcsname

\title{Universal upper critical field of unconventional  superconductors} 
\author{V.N. Zavaritsky$^{1,2}$, V.V. Kabanov $^{3}$,  and A.S.
Alexandrov$^{1}$\\ 
$^{1}$Department of Physics, Loughborough University, Loughborough LE11
3TU, United Kingdom;\\
$^{2}$Kapitza Institute for Physical Problems, 2 Kosygina
Str., 117334 Moscow, Russia;\\
$^{3}$Josef Stefan Institute 1001, Ljubljana, Slovenia.} 

\date{\today}

\maketitle

\begin{abstract}
The resistive upper critical field, $H_{c2}(T)$ of  cuprates,
superconducting spin-ladders, and organic (TMTSF)$_2$X systems is shown to follow a
universal nonlinear dependence $H_{c2} \propto
(T_c-T)^{3/2}$ in a wide range near $ T_c $, while 
its low-temperature behaviour depends on the
chemical formula and sample quality. $H_{c2}(T)$ is
ascribed to the Bose-Einstein condensation field of preformed pairs.  
The universality originates from the scaling arguments. Exceeding
the Pauli paramagnetic limit is explained. Controversy in the
determination of $H_{c2}(T)$ from the  kinetic and thermodynamic
measurements is resolved in the framework of the charged Bose-gas 
model with impurity scattering.
\end{abstract}
 
\vskip2pc]

\narrowtext

The upper critical field is one of the fundamental characteristics of
type II superconductors. For sufficiently high field,
superconductivity is destroyed and the field is uniform in a bulk
sample. Continuously decreasing the field superconducting regions
begin to nucleate spontaneously at a certain filed $B=H_{c2}(T)$.
 In the regions where the nucleation occurs, superconductivity is just
beginning to appear, so that the density of supercarriers,
$n_s=|\psi({\bf r})|^2$ is small. Hence, the phenomenological
Landau-Ginsburg (LG) (or the microscopic Gor'kov) equation for the order
parameter $\psi({\bf r})$  can be linearized to give
\begin{equation}
{1\over{2m}}
(\nabla-2ie{\bf A}({\bf r}))^{2}\psi({\bf r})=\alpha \psi({\bf r}),
\end{equation} 
where 
$\hbar=c=k_B=1$. 
$H_{c2}(T)$ allows for a direct
 measurement of the most fundamental parameter,
the superconducting coherence length, $\xi(T)$, because
$H_{c2}=\phi_0/2\pi \xi(T)^2$ ($\phi_0$ is the flux quantum) 
\cite{degen}. Solving Eq.(1), one obtains the linear 
 $H_{c2} (T)=-m\alpha/e$ near
 $T_c$ \cite{tc} with $\alpha \propto T-T_c$ in the
 Landau theory of the  second-order phase transitions. At
 zero temperature $H_{c2}(0)$ is normally below the
 Clogston-Chandrasekhar \cite{clog} or the Pauli pair-breaking limit given
 by $H_p \simeq 1.84 T_c$ (in Tesla) for the singlet pairing. The
 limit can be exceeded due to the spin-orbit coupling \cite{hoe}, or
 triplet pairing, but in any case $H_{c2}(0)$ remains finite in the 
framework of the BCS theory.
The mean-field BCS approach, Eq.(1), is
 applied if $\xi(0) >> (6/n\pi)^{1/3}$, where $n$ is the carrier
 density. Hence, irrespective to the Pauli pair-breaking limit, the zero 
temperature value of the (BCS) upper critical
 field should be  much less than $\phi_0 (\pi n/6)^{2/3}/(2\pi)$, which is
 about 200 Tesla for a typical carrier density  in
 novel superconductors ($n \simeq 10^{21} cm^{-3}$). 

In cuprates \cite{gen,mac,oso,alezav,and,car,zn-ybco,boe,zav,gan,gan2,zav2}, 
spin-ladders \cite{spin} and organic
superconductors \cite{org} high magnetic field studies revealed a
non-BCS upward curvature of resistive $H_{c2}(T)$.
When  measurements were
performed on low-T$_c$ unconventional superconductors 
\cite{mac,oso,zn-ybco,spin,org}, the Pauli limit
was exceeded by several times. A non-linear temperature
dependence in the vicinity of  T$_{c}$ was
unambiguously observed in a few samples \cite{alezav,zn-ybco,gan,gan2,zav2}. 
This  strong  departure
from the canonical BCS behaviour led some
authors \cite{car,wen,mor,luo} to conclude, that the abrupt resistive
transition in applied fields is not a normal-superconductor transition
at $H_{c2}$. Indeed, the  thermodynamic determination of $H_{c2}$
\cite{car,luo,jun,jun2}, and  anomalous diamagnetism above the resistive
transition \cite{jun3,wen} seem to justify such a
conclusion.  Thermodynamically determined $H_{c2}$ appears to be linear in the
vicinity of $T_{c}$, and much higher than the
resistive $H_{c2}$,   in some cases
\cite{wen,luo} exceeding well not only the Pauli limit, but even the ultimate 
'BCS' limit mentioned above. 

The apparent controversy in  different
determinations of $H_{c2}$ needs to be addressed beyond the mean-field
approach, Eq.(1). Unconventional
superconductors could be in the 'bosonic' limit of preformed
real-space pairs, so their resistive $H_{c2}$ is actually
a critical field of the Bose-Einstein condensation of charged
bosons, as proposed by one of us
\cite{ale}. The calculations \cite{alekab} carried out for the heat
capacity of an ideal charged Bose-gas in a magnetic field 
%led  to the conclusion that 
revealed a remarkable difference between the resistive $H_{c2}$ and the thermodynamically
determined one. 
While any magnetic
field destroys the condensate of ideal bosons, it hardly shifts the
specific heat anomaly. 

In this Letter, we present a comprehensive scaling 
of resistive $H_{c2}$ measurements  in a great variety of unconventional superconductors.  
A universal non-BCS  temperature
dependence 
is found in the vicinity of $T_c$ 
while deviations from the universality  are observed at low temperatures. 
We describe these results in the framework of a microscopic 
model of charged bosons
scattered off impurities. Different from the ideal Bose-gas this
model predicts $two$ anomalies in  the specific heat.
The lower  temperature anomaly traces the resistive transition in a
magnetic filed, but the higher one is hardly shifted even by
a  high magnetic field, as observed 
\cite{jun,jun2}. Based on the microscopic model we argue that the state above
the resistive $H_{c2}(T)$ of unconventional superconductors is the
$normal$ state of preformed pairs.

In the bosonic superconductor the mean-field LG equation, Eq.(1) is replaced 
by the microscopic Schr\"odinger equation for
the condensate wave function \cite{ale},
\begin{equation}
[\hat{U}_{sc}-(\nabla-2ie{\bf A}({\bf r}))^{2}/(2m)]\psi({\bf r})=\mu \psi({\bf r}),
\end{equation} 
where $\hat{U}_{sc}$ is the scattering potential due to impurities and 
phonons, or the self-energy operator due to
interparticle hard-core and long-range correlations \cite{alebeerkab}, 
and $\mu$ is the chemical potential. 
Different from the mean-field Eq.(1), it takes fully into account both thermal 
and quantum fluctuations, but 
does not allow for a direct determination of $H_{c2}$. When $H_{c2}$ is 
defined as the field
where the first non-zero extended solution of Eq.(2) appears, the equation 
yields a position of the 
chemical potential at the mobility edge $\mu=E_{c}$, rather than $H_{c2}$ 
itself. Then the upper critical field is 
found using the total number of extended bosons $n_b$ above the mobility edge,
\begin{equation}
\int_{E_{c}}^{\infty}f(\varepsilon)N(\varepsilon,H_{c2})d\varepsilon
=n_b (T),
\end{equation}
where $N(\varepsilon,B)$ is the density of states (DOS) of the Hamiltonian, Eq.(2), and $f(\varepsilon)=1/[exp((\varepsilon-\mu)/T-1]$ is the
Bose-Eistein distribution. In the general case 
$n_{b}(T)$ depends on temperature due to a partial localization of bosons in 
the random potential.

 Applying simple scaling arguments \cite{ale} the positive curvature of 
$H_{c2}(T)$ near $T_{c}$ and its divergent behaviour at low temperatures
follow   from Eq.(3).
  The number of bosons
at the lowest Landau level ($n=0$) is proportional to the temperature and 
 DOS near
the mobility edge, $N_{0}\propto B/\sqrt{\Gamma_0(B)}$. The collision 
broadening
of the Landau level is also proportional to the same DOS
$\Gamma_0(B)\propto B/\sqrt{\Gamma_0(B)}$. Hence, $\Gamma_0(B)\sim B^{2/3}$ 
and therefore the
number of  bosons at the lowest level is proportional to $T B^{2/3}$. 
The singularity of all upper levels' DOS is integrated out in 
Eq.(3) so that 
one can neglect their quantization using the zero field density of states for 
the levels with $n \geq 1$. 
Equating the number of bosons with $n=0$ and $n_b(T)-n_b(T_c)(T/T_c)^{3/2}$ 
(the total number minus the number of thermally excited bosons with $n \geq 1$)  
yields 
\begin{equation}
H_{c2}(T)=H_0 [n_b(T)/(t n_b(T_c))-t^{1/2}]^{3/2},
\end{equation}
where $t=T/T_c$. The scaling constant 
$H_0$ depends on the scattering mechanism, $H_0= \phi_0/2\pi \xi_0^2$, with 
the characteristic (coherence) length
$ \xi_0 \simeq (l/n_{b}(T_c))^{1/4}$. Here $l$ is the zero-field mean-free 
path of the low energy bosons. 
One obtains the parameter-free $H_{c2}(T)\propto (1-t)^{3/2}$ using Eq.(4) in the vicinity of $T_c$, 
but the low-temperature 
behaviour depends on the particular scattering mechanism, and  the
detailed structure of the density of localized states. 
As suggested by the normal state Hall measurements in cuprates \cite{alebra} 
$n_b(T)$ can be parameterized
 as $n_b(T)=n_b(0)+constant\times T$, so that $H_{c2}(T)$ is described 
by a single-parameter expression as
\begin{equation}
H_{c2}(T)=H_0 [b(1-t)/t +1-t^{1/2}]^{3/2}.
\end{equation}
Parameter $b$ is proportional to the number
 of delocalised bosons at zero temperature.
We expect that this expression  applies to the whole temperature
 range except 
ultra-low temperatures, where  the Fermi Golden-rule in the scaling fails 
\cite{alebeerkab}.
Exceeding the Pauli pair-breaking limit readily follows from the fact, that 
the singlet-pair binding energy
 is 
related to the normal-state pseudogap temperature $T^*$, rather than
 to $T_c$  \cite{alemot}. $T^*$ is
higher than $T_c$  in bosonic superconductors, and cuprates.

The universal  scaling of  $H_{c2}$ near $T_c$ is confirmed by the resistive 
measurements
of the upper critical field of many cuprates, spin-ladders, 
and organic superconductors, as shown in Fig.1A. All data reveal the
 universal $(1-t)^{3/2}$ 
behaviour in a wide temperature region as can be seen in the inset to Fig.1A. Deviations from
this law, observed in a few cuprates in a close vicinity of $T_c$ were explained in Ref. \cite{alezav}.
The low-temperature behaviour of $H_{c2}(T)/H_{0}$ is not universal, 
but well described using Eq. (5) with a single  fitting parameter,
$b$. This is close to 1 in  high quality
 cuprates with a very narrow resistive transition
\cite{alezav,gan,zav2}. It naturally becomes rather small in 
overdoped cuprates where the randomness is more essential, so almost all bosons 
are localized 
(at least in one dimension) at zero temperature. It becomes even smaller in 
organic superconductors, 
which might be related  to the magnetic field induced dimensional crossover 
\cite{leb} at low temperatures. The scaling parameter $H_0$ increases with increasing $T_c$, Fig.1B.
This is because mean-free path $l$ decreases with doping, while  the density of carriers increases, so that the coherence length $\xi_0$ becomes smaller in the cuprates with a higher $T_c$.

Calculations of the specific heat require the analytical DOS, 
$N(\varepsilon,B)$ of a particle in 
the random potential and in the magnetic field. The above scaling suggests 
that $H_{c2}(T)$ is not  sensitive to
a particular choice of the scattering mechanism and approximation, at 
least in a wide vicinity of $T_c$.
Hence, one can use the canonical non-crossing approximation for the 
single-particle self-energy,
\begin{equation}
\Sigma_\nu (\varepsilon)=\sum_{\nu'} 
\frac{\Gamma_{\nu,\nu'}}{\varepsilon-\varepsilon_{\nu'}-\Sigma_{\nu'}(\varepsilon)}
\end{equation}
with a particular scattering matrix element squared
$\Gamma_{\nu,\nu'}=\Gamma \delta_{n,n'}$, $\nu \equiv (n,p_x,p_z)$ are the 
quantum numbers of the Landau problem. This allows us to obtain an analytical 
result for the DOS,
$N(\varepsilon,B)=\Gamma^{-1}\sum_{n} \Im \Sigma_n (\varepsilon)$ as
\begin{eqnarray}
N(\varepsilon,B)&=&
\frac{eB}{4\pi^{2}}{\sqrt\frac{6m}{\Gamma_0}}
\sum_n \Bigg[\left({\tilde{\varepsilon_n}^{3}}+{1\over{2}}+\sqrt
{{\tilde{\varepsilon_n}^{3}}+
{1\over{4}}}\right)^{1/3}\cr&-&\left({\tilde{\varepsilon_n}^{3}}+
{1\over{2}}-\sqrt
{{\tilde{\varepsilon_n}^{3}}+{1\over{4}}}\right)^{1/3}\Bigg],
\end{eqnarray}
with the mobility edge at
$E_{c}=eB/m-3\Gamma_{0}/2^{2/3}$.
Here $\Gamma_{0}=0.5(2\Gamma eB\sqrt{m}/\pi)^{2/3}$ is the 
collision broadening of the lowest Landau level, % and 
$\tilde\varepsilon_n=[\varepsilon-2eB (n+1/2)/m]/3\Gamma_{0}$. 

$H_{c2}(T)$ calculated with
the analytical DOS, Eq.(7) is almost the same as $H_{c2}$ in Eq.(5). 
The specific heat  coefficient 
$C(T,B)/T=d[\int d\varepsilon N(\varepsilon,B)\varepsilon f(\varepsilon)]/TdT$ 
calculated with the same DOS and with $\mu$
determined from $n_b=\int d\varepsilon N(\varepsilon, B)f(\varepsilon)$ is shown in
Fig.2a. The broad maximum at $T \simeq T_c$ is practically the same as in the
ideal Bose gas without scattering \cite{alebeerkab}. It barely shifts in the
magnetic field. However, there is the other anomaly at lower temperatures,
which is absent in the ideal gas. It shifts with the magnetic field, tracing
 the resistive transition, as clearly seen  from the difference
between the specific heat in a field and the zero-field curve, Fig. 2b. The
specific heat, Fig. 2, is  in striking resemblance to the Geneva group's
experiments on DyBa$_2$Cu$_3$O$_7$ (Fig. 4 and 6 in Ref.\cite{jun2}) and on
YBa$_2$Cu$_3$O$_7$ (Fig. 1 and 2 in Ref.\cite{jun}), where both anomalies
were observed. 
 
Within our model,  when the magnetic field is applied, it  hardly changes 
the temperature 
dependence of the chemical potential near  $T_c$ since the 
energy spectrum of thermally excited bosons remains practically unchanged. That is 
because their characteristic energy (of the order of $T_c$) remains enormous 
compared with the magnetic energy of the order of $2eB/m$. In contrast, the 
energy spectrum of the low energy bosons is
strongly perturbed even by a weak magnetic field. As a result the chemical 
potential 'touches' the band edge at lower temperatures, while having almost 
the same 'kink'-like temperature dependence around $T_c$ as in zero field.
While the lower anomaly corresponds to the true long-range order,  the higher 
one is just a trace-'memory'  of the zero-field transition. Hence, our microscopic 
consideration shows  that the genuine
 phase transition into the superconducting state is 
related to the resistive transition and to the lower specific heat anomaly. 
The broad higher anomaly is the normal state feature of the bosonic system
in the external magnetic field. Different
 from the BCS superconductor these two anomalies are well separated in
 the bosonic superconductor at any field except zero one.
Hence, the resistive $H_{c2}$ is the genuine upper critical field, while the field $H^*$ determined 
thermodynamically from the higher
anomaly of the specific heat, Fig.~2b, is a pseudo-critical field, unrelated 
directly to the long-range off-diagonal superconducting order. 
The absence of  significant superconducting
fluctuations in the resistivity of the highest quality samples 
\cite{gan,mac,oso,zn-ybco,zav} in a wide field interval
between
the resistive $H_{c2}(T)$ and  $H^*$ further justifies the conclusion. A
weak diamagnetism observed in a few cuprates
above the resistive $H_{c2}(T)$ curve \cite{jun3,wen},
 was  explained as the $normal$ state Landau diamagnetism of 
preformed pairs in the framework of the same microscopic model of charged 
bosons \cite{mag}. 

Our conclusions are at variance with some others \cite{mor}, which claim
that strongly anisotropic Bi-cuprates remain in the superconducting state well above 
the resistive $H_{c2}(T)$. However, thorough analysis \cite{condmat} of the data 
used by \cite{mor} to support that claim reveals significant contribution from 
extrinsic effects. These are responsible for the apparent contradiction between the
results of \cite{mor} and those of the predecessors \cite{alezav,zav,zav2}.
In particular, as shown in \cite{condmat}, the unusual shape of $\rho_{ab}(H)$ \cite{mor} could
result from the current redistribution in a defective crystal while the Joule heating  is
likely to be responsible for the non-Ohmic resistance observed in \cite{mor}. Moreover, as shown in Fig.~3,
when the routine procedure for the resistive $H_{c2}$ evaluation
\cite{alezav} is applied to reliable in-plane and out-of-plane data
obtained on the $same$ samples \cite{and,zha}, very similar values of $H_{c2}(T)$
are  obtained from $\rho_c$ and $\rho_{ab}$ \cite{condmat}. 
This puts into question the last argument of the authors of Ref. \cite{mor} 
who claim that while $\rho_c$ is a measure of the
interplane tunneling, only the in-plane resistivity %, $\rho_{ab}$
represents a true normal state and should be used in the determination of
$H_{c2}$. It is appropriate to mention here that
according to the analysis in Fig.1,
significant features of the resistive $H_{c2}$ appear to
be robust with respect to the particular component of the resistivity used for its evaluation.
Finally, the observation \cite{zav} of the negative c-axis magnetoresistance above $T_c$ invalidates 
the main claim of Ref. \cite{mor} that it is a signature of the superconducting state. 

In conclusion, we have scaled the  magnetotransport
measurements in many novel superconductors.  The unusual upper
critical field $H_{c2}(T)$ has been ascribed to the
Bose-Einstein condensation field of preformed pairs. We have
introduced a charge Bose-gas model with a particular choice of the
scattering potential allowing for the analytical DOS in the magnetic
field. In contrast to an ideal Bose-gas model and the BCS theory, this model 
describes well
resistive  $H_{c2}(T)$ and predicts two anomalies in the specific
heat. We have shown that  the genuine
 phase transition into the superconducting state is 
related to the resistive transition and to the lower specific heat anomaly, 
while the 
higher one is the normal state feature of the bosonic system
in the external magnetic field. Our approach is compatible with a wealth
of various experimental observations, the normal pseudogap
and the absence of the Hebel-Slichter peak being only a few of them \cite{alemot}.

The authors
acknowledge valuable discussions with A. F. Andreev,
V. F. Gantmakher, L. P. Gor'kov, A. Junod,
W. Y. Liang, J. W. Loram, V. V. Moshchalkov,  M. Springford, and
G. M. Zhao.  The financial support of the EPSRC (Ref.:  R/46977) and the Leverhulme Trust (Ref.: F/00261/H)
is gratefully acknowledged.

\centerline{{\bf Figure Captures}}

Fig. 1. {\bf A}: Resistive upper critical field (evaluated at 50\% of the
transition) of  electron/hole doped cuprates, spin-ladders and organic superconductors
scaled according to Eq.(5). Parameter b is 1 (solid line), 0.02
(dashed-dotted line), 0.0012 (dotted line), and 0 (dashed line).  Inset shows the
universal scaling of the same data near $T_c$. {\bf B}: $H_0$ versus $T_c/T_c^{opt}$, 
where $T_c^{opt}$ is the critical temperature of the 
optimally doped material (shown in brackets). Lines are the guide for eyes. 
The right-hand part of {\bf B} shows $H_0$ for the compounds where $T_c^{opt}$ is
unknown.

Fig.2. Temperature dependence of the specific heat $C(H,T)/T$ 
(in units of $2nk_B/[\pi^{1/2}\zeta(3/2)T_c]$) of charged
Bose-gas scattered off impurities for several fields indicated in the figure ($\omega_H=2eB/m$). 
Fig.~2b: Likewise \cite{jun,jun2,jun3} shows $C(H,T)-C(0,T)$ and reveals two anomalies: 
the lowest traces the resistive transition while the
highest, $H^*$, is the normal state feature.

Fig.3. $H_{c2}(T)$ obtained from $\rho_{ab}$ (solid lines) and $\rho_c$ (dashed lines)
of the same sample \cite{and,zha}. 

 
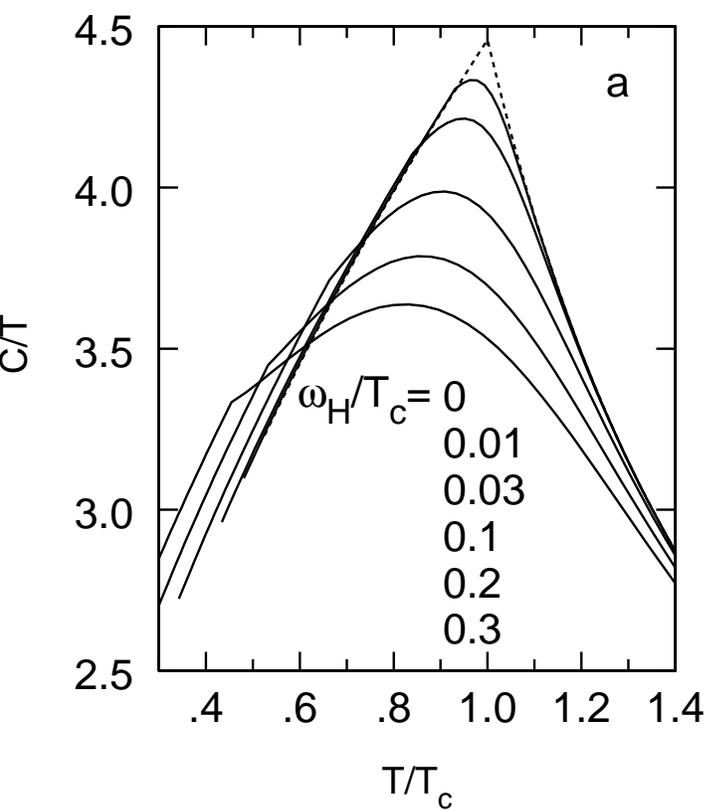
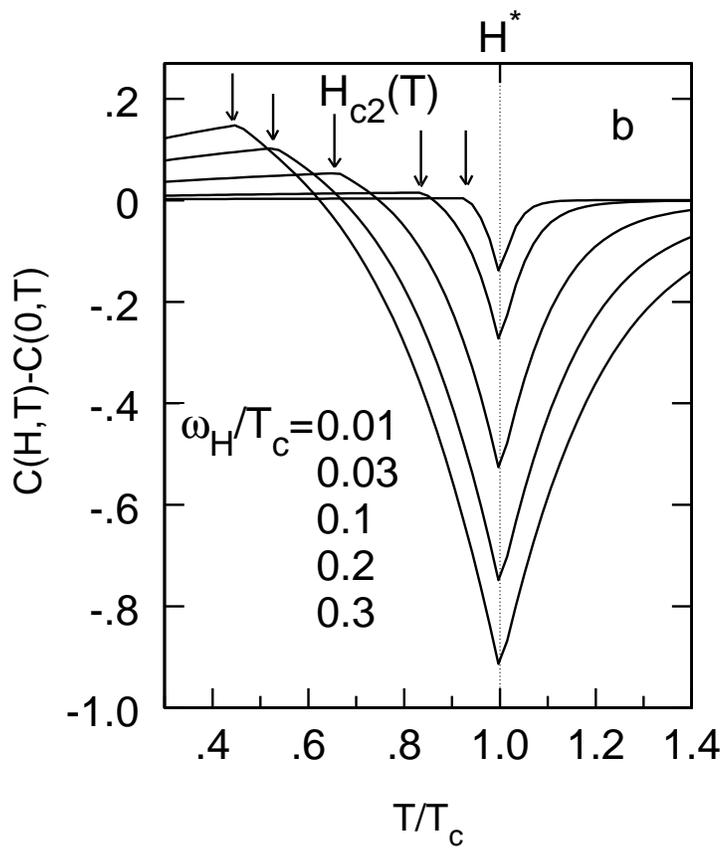
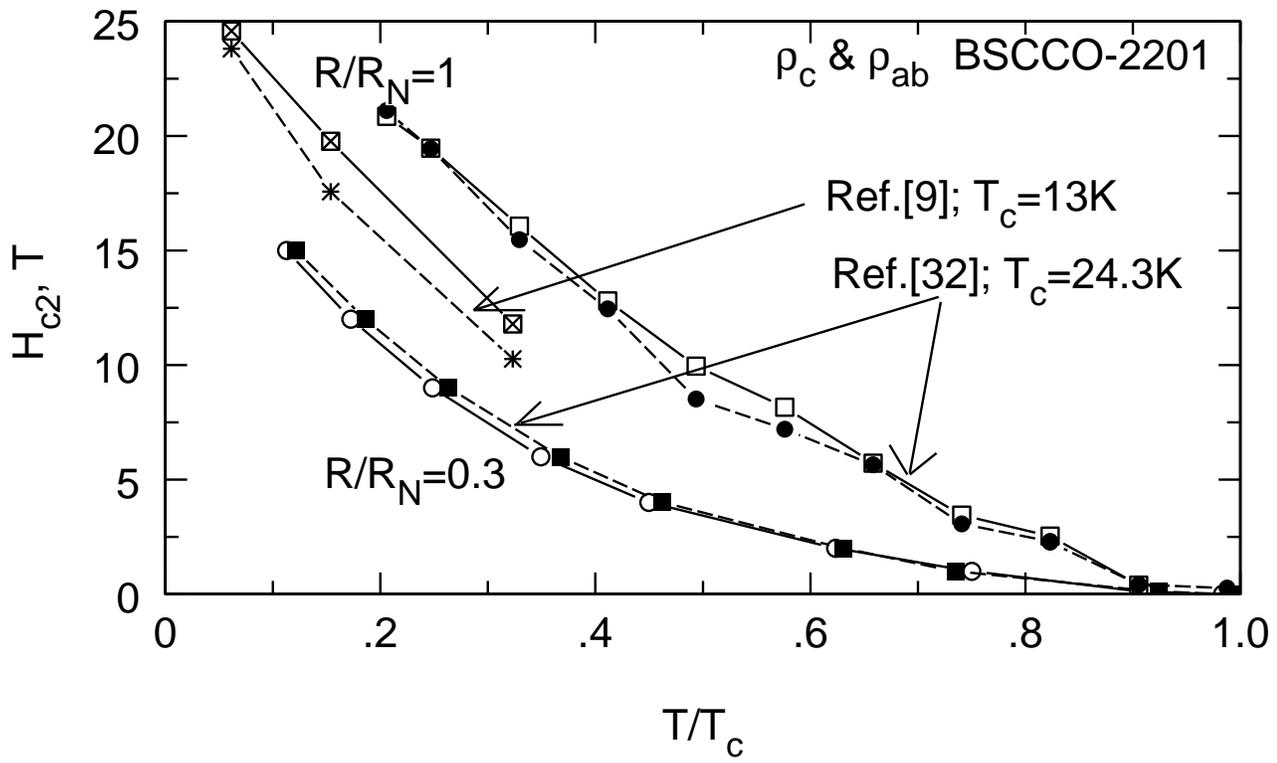
\end{document}